\documentclass[twocolumn,showpacs,preprintnumbers,%
amsmath,amssymb,aps,floatfix]{revtex4}

\usepackage{graphicx}
\usepackage{dcolumn}
\usepackage{bm}

\def\vek#1{ {\rm \bf  #1} }
\def \vekk#1{ {\bm #1} }

\begin{document}

\title{Quasistatic limit of the strong-field approximation describing 
       atoms and molecules in intense laser fields}  

\author{Yulian V. Vanne}
\author{Alejandro Saenz}%
\affiliation{%
AG Moderne Optik, Institut f\"ur Physik, Humboldt-Universit\"at
         zu Berlin, 
         Hausvogteiplatz 5-7, D\,--\,10\,117 Berlin, Germany}%

\date{\today}%

\begin{abstract}

The quasistatic limit of the velocity-gauge strong-field approximation  
describing the ionization rate of atomic or molecular systems exposed to 
linear polarized laser fields is derived. It is shown that in the 
low-frequency limit the ionization rate is proportional to the laser 
frequency, if a Coulombic long-range interaction is present. An expression 
for the corresponding proportionality coefficient is given. Since neither 
the saddle-point approximation nor the one of a small kinetic momentum is 
used in the derivation, the obtained expression represents the exact 
asymptotic limit. This result is used to propose a Coulomb 
correction factor.  Finally, the applicability of the 
found asymptotic expression for non-vanishing laser frequencies is 
investigated.  

\end{abstract}

\pacs{32.80.Rm, 33.80.Rv}

\maketitle

\section{\label{sec:Intro}Introduction} 

Keldysh-Faisal-Reiss (KFR) theories are very popular to describe 
nonresonant multiphoton ionization of atoms and molecules in intense laser 
fields (see, e.\,g.,~\cite{sfa:beck01,sfa:beck05,sfa:kjel06,sfa:milo06} and
references therein). 
While the principle concept of ignoring the effect of the laser field in 
the initial state and the interaction of the ionized electron with the 
remaining atomic system in the final state is common to all KFR 
approximations, the approaches differ in the details of their formulation. 
Whereas the length gauge was used in the original work of 
Keldysh~\cite{sfa:keld65}, the velocity gauge was invoked by 
Reiss~\cite{sfa:reis80} and Faisal~\cite{sfa:fais73}. Historically, the 
velocity-gauge variant of KFR theory is also known as strong-field 
approximation (SFA)~\cite{sfa:reis92}. Although this terminology is not 
consequently adopted nowadays, this meaning of SFA is used in the present 
work that discusses exclusively the velocity-gauge variant of KFR. 

A natural test of KFR theories is a comparison of its prediction 
in the tunneling limit with the corresponding one of quasistatic 
theories~\cite{sfa:pere66,sfa:ammo86}. The tunneling limit of the 
length-gauge version of KFR was considered already by Keldysh for the explicit 
example of a hydrogen atom. In the derivation he employed, however, 
two additional approximations: the one of a small kinetic momentum 
and the saddle-point method (SPM). In his detailed 
work~\cite{sfa:reis80} about SFA theory  
Reiss has also used the SPM to obtain an approximation for the generalized 
Bessel functions ('asymptotic approximation') that are required for the 
calculation of the transition amplitude. As in Keldysh' work he also 
adopted an expansion assuming a small kinetic momentum in order to derive 
the differential ionization rate in the tunneling limit. The algebraically 
cumbersome form of the in~\cite{sfa:reis80} derived 'asymptotic approximation' 
has motivated the development of a simpler form appropriate for the 
tunneling regime. This approximation~\cite{sfa:reis03} is again 
based on the SPM and is referred to as 'tunneling 1 approximation'. 
It was, however, recently criticized by Bauer~\cite{sfa:baue05c} who has 
shown numerically that the 'tunneling 1 approximation' is much worse than the 
'asymptotic approximation' in the case of large kinetic momenta.  
Since neither~\cite{sfa:reis80} nor~\cite{sfa:reis03} contain 
explicit asymptotic expressions in the limit of 
vanishing laser frequency ($\omega\rightarrow 0$), a systematic study of 
the quasistatic limit on their basis is almost impossible. 
Recently, the different versions of KFR were numerically  
compared with quasi-static theories  for the 
hydrogen atom in~\cite{sfa:baue06}. %
It was found that SFA significantly underestimates the ionization rate, 
especially in the limit $\omega\rightarrow 0$ or for very strong fields. 
Since both limits can be well described with quasistatic theories,
a comparison of them with the corresponding limit of SFA can 
provide an insight into the reasons for such a disagreement. The derivation
and analysis of such an asymptotic expression for the SFA ionization rate 
is the main motivation of this work. As is shown below, the SFA rate does 
not converge to the tunneling result, if long-range Coulomb interactions 
are present. 

The asymptotic limit of SFA for $\omega\rightarrow 0$ is also very 
interesting, because it may be used for deriving a Coulomb correction 
factor by comparing this limiting expression with the one of quasistatic 
theories. Rescaling the SFA rate (for $\omega\neq 0$) in such a way that 
it agrees with the quasistatic limit for $\omega\rightarrow 0$ is supposed 
to correct SFA for the otherwise neglected long-range Coulomb interaction 
between the ionized electron and the remaining ion. Such a Coulomb 
correction factor was proposed by Becker and Faisal 
(see \cite{sfa:beck05} and references therein) and is extensively used 
in their atomic and molecular SFA calculations\cite{sfa:beck01,sfa:beck05}. 
Note, this correction 
factor is in fact very large and can amount to almost three orders of 
magnitude for atomic hydrogen and standard parameters of intense 
femtosecond lasers. Although it is emphasized   
in \cite{sfa:beck05} that the low-frequency limit of SFA converges to 
the tunneling result, this is only shown for the case of short-range 
interactions. As is demonstrated in the present work, the correct 
asymptotic limit of SFA in the presence of long-range Coulomb 
interactions differs from the short-range case even {\it qualitatively}, 
since it is proportional to $\omega$, but $\omega$ independent for 
short-range potentials. Therefore, the present work also allows to 
directly derive an asymptotically correct Coulomb correction factor 
for SFA.    

The present paper is organized the following way. After a brief description 
of the ionization rate within SFA in which the basic formulas and 
notations are introduced (Sec.~\ref{subsec:rate}), an expression is derived 
in Sec.~\ref{subsec:Efc} that is numerically very convenient for the 
calculation of generalized Bessel functions and thus the SFA in the 
quasistatic limit. In Sec.~\ref{subsec:Bessel} an exact asymptotic formula 
is derived for the generalized Bessel functions in the limit 
$\omega\rightarrow 0$. In this derivation neither the SPM nor any other 
approximation beyond the ones inherent to SFA are used and it is 
demonstrated that the SPM yields wrong results for weakly bound systems 
or very intense fields. In Sec.~\ref{subsec:QSFA} 
two simplifications are introduced that in contrast to the SPM or 
small-momentum approximation 
are universally justified in the limit $\omega\rightarrow 0$. This allows 
to derive an exact analytical expression of the quasistatic limit of the SFA 
in the presence of long-range interactions that we name {\it QSFA}. 
In Sec.~\ref{sec:hydrogen} the QSFA is discussed for the example  
of atomic hydrogen. After a derivation of the parameters specific to 
the considered atomic system in Sec.~\ref{subsec:Coeff}, 
the rate obtained in the weak-field limit is 
discussed and compared to tunneling models in Sec.~\ref{subsec:weak}. 
Based on this comparison, a Coulomb correction factor is derived for 
SFA and compared to an earlier proposed one. The range of validity of 
QSFA for standard laser frequencies is explored in 
Sec.~\ref{subsec:validity} where also a correction is proposed that is 
explicitly given for the 1S state of hydrogenic atoms. 
The findings of this work are summarized in Sec.~\ref{sec:conclusion}.

\section{\label{sec:Theory}Theory} 

\subsection{\label{subsec:rate}Ionization rate}

In the single-active-electron approximation, we consider the direct transition 
of an electron from the initial bound state $\Psi_0$ to a continuum 
state $\Psi_{\vek{p}}$ due to the linear polarized laser 
field $\vek{F}(t) = \vek{F} \cos\omega t$ with period $T = 2\pi/ \omega$. 
The total ionization rate is given in the SFA by
\begin{eqnarray}
\nonumber
   W_{\rm SFA} &=&   (2\pi)^{-2}  \sum_{N \ge N_{0}}\, p_N 
                     \left( \frac{p^2_N}{2} + E_{\rm b} \right)^2 \\
               &\times&
                     \int\limits  d \hat{\vek{p}}\,| L(p_N \hat{\vek{p}})|^2\, 
                     |\tilde{\Psi}_0(p_N, \hat{\vek{p}})|^2
\label{W}
\end{eqnarray}
with $\kappa = \sqrt{2 E_b}$ where $E_b$ is the binding energy of the initial 
bound state $\Psi_0$ with its Fourier transform $\tilde{\Psi}_0$. The number 
of absorbed photons $N$ satisfies $N\ge N_0 =  (E_{\rm b} +  U_p)/\omega$ 
where $U_p = F^2/(4\omega^2)$ is the electron's quiver (ponderomotive) energy 
due to the laser field. Finally, $p_N = \sqrt{ 2(N\omega - E_{\rm b} - U_p)}$ 
is the momentum in the final state for an $N$ photon transition.  
The function $L(\vek{p})$ is defined as
\begin{equation}
   L(\vek{p}) = \frac{1}{T}\int\limits_{0}^{T} e^{i S_{\vek{p}}(t)} d t
\label{Lp}
\end{equation}
where $S(t)$ is given with the aid of the mechanical momentum of the 
electron, $\vekk{\pi}(t) = \vek{p} + (\vek{F}/\omega) \sin\omega t$, as
\begin{equation}
   S_{\vek{p}}(t) =  \int\limits_{0}^{t} d t' \left[ E_{\rm b} +  
                     \frac{1}{2}\vekk{\pi}^2(t') \right]   \quad .
\label{St}
\end{equation}
For $\vek{p} = p_N\hat{\vek{p}}$ the function $L(\vek{p})$ can simply be 
expressed using the generalized Bessel functions (we use for them Reiss' 
definition which differs slightly from the one of Faisal) as
\begin{equation}
   L(p_N\hat{\vek{p}}) =  (-i)^N e^{\xi i} J_N( \xi, -z/2 )
\label{LJN}
\end{equation}
where $\xi = p_N F (\hat{\vek{p}}\cdot\hat{\vek{F}})/\omega^2$ and 
$z = U_p/\omega$. In the high-frequency and low-intensity (so called 
multiphoton) regime the generalized Bessel functions can be 
very efficiently calculated using an expansion over products of ordinary 
Bessel functions, 
\begin{equation}
   J_N(a,b) = \sum_{m=-\infty}^{\infty} J_{N-2m}(a) J_{m}(b) \quad ,
\label{JNexp}
\end{equation}
where only a few terms are required to yield high accuracy.

Consider now the quasistatic limit defined by $\omega \rightarrow 0$. 
Introducing the Keldysh parameter 
$\gamma = \kappa\omega/F$, the (inverse) field parameter 
$\tau=\kappa^3/F$, and the variables
\begin{equation}
   q_N = p_N/\kappa \qquad {\rm and} \qquad 
                           \zeta=\hat{\vek{p}}\cdot\hat{\vek{F}},
\label{qNzeta}
\end{equation}
one finds
\begin{equation}
   \xi = \frac{\tau q_N \zeta}{\gamma^2},
         \quad 
   z   = \frac{\tau}{4\gamma^3},
         \quad 
   N_0 = \frac{\tau}{4\gamma^3}(1 + 2 \gamma^2).
\end{equation}
The condition $\omega \rightarrow 0$ leads to  
$\gamma \rightarrow 0$, whereas parameter $\tau$ is $\omega$ independent 
and thus unaffected. Since the numerical values of $q_N$ and $\zeta$ are 
usually of the order of one, both arguments and the index of the 
generalized Bessel function in (\ref{LJN}) approach to infinity. %
In this case it is very problematic to use (\ref{JNexp}) for numerical 
calculations, since very many terms are required and their amplitudes are 
much larger than the final result. This can lead to large cancellation errors. 
In the next subsection we solve this problem by a transformation of the 
integral~(\ref{Lp}) to a form that is more convenient for numerical 
calculations.

\subsection{\label{subsec:Efc} Efficient calculation in the quasistatic limit}

A very efficient way for the numerical computation of $L(p_N\hat{\vek{p}})$
in the tunneling regime is possible by means of performing the integration 
through the saddle points. Introduction of the new complex variable 
$u = \sin\omega t$ allows to rewrite (\ref{Lp}) as
\begin{equation}
  L = \oint\limits_{C_{\rm c}} {\cal F}(u) d u
\label{LpCom}
\end{equation}
where
\begin{eqnarray}
   {\cal F}(u) &=& \frac{e^{i S(u)}}{2\pi f(u)}, \\
    S(u) &=& \frac{\tau}{2\gamma^3}\int\limits_{C_u} 
            \frac{v^2 +2\gamma q_N \zeta v + \gamma^2(1 + q^2_N)}{f(v)} d v ,\\
\label{Su}
   f(u) &=& {\rm Sign}[{\rm Im}(u)] \sqrt{1-u^2}.
\label{fu}
\end{eqnarray}
The closed contour $C_{\rm c}$ in Eq.~(\ref{LpCom}) encloses the branch cut 
$[-1,1]$ of the functions $f(u)$, $S(u)$, and ${\cal F}(u)$. The path of 
integration $C_u$ in  Eq.~(\ref{Su}) specifies the path around the branch 
cut starting at $v= i 0^{+}$ and terminating at $v = u$. Since $S(u)$ is 
a multivalued function, we have selected the branch cut along the negative 
imaginary axis. Nevertheless, function ${\cal F}(u)$ (as well as $f(u)$ ) is
analytical in the whole complex plane except its branch cut $[-1,1]$.

There exist two saddle points $u_{\pm}$ of $S(u)$ in the complex plane $u$ 
defined by $S'(u_{\pm})=0$ and given explicitly by
\begin{equation}
   u_{\pm} = \gamma\rho\, (-\chi \pm  i)
\label{upm}
\end{equation}
with
\begin{equation}
   \rho = \sqrt{1 + q_N^2(1 - \zeta^2)}\ge 1,\quad \chi 
        = q_N \zeta/\rho \le q_N   \quad .
\label{rhochi}
\end{equation} 
We introduce the straight contours $C_{\pm}$ that go through the saddle 
points $u_{\pm}$ and are given parametrically as
\begin{equation}
   u(x, Q_{\pm}) = u_{\pm} + x Q_{\pm}, \quad -\infty < x <\infty
\label{upar}
\end{equation}
starting at $x \rightarrow - \infty$. The values of $Q_{\pm}$ are chosen in 
such a way that the contours $C_{\pm}$ are passing through the steepest 
descent, i.\,e.\ as
\begin{equation}
   Q_{\pm} = \sqrt{\frac{2i}{S''(u_{\pm})}} 
\end{equation}
where the argument of $Q_{\pm}$ satisfies $- \pi/4 < \arg Q_{\pm} < \pi/4$.

As $|u|\rightarrow \infty$, the function ${\cal F}(u)$ decays exponentially 
to 0 for $|\arg u| < \pi/4$. This allows to transform the contour integral 
(\ref{LpCom}) as 
\begin{equation}
   L = \oint\limits_{C_{+}} {\cal F}(u) d u - 
            \oint\limits_{C_{-}} {\cal F}(u) d u 
     = L_{+} - L_{-} 
\label{Lsum}
\end{equation}
where the integrals $L_{\pm}$ can be calculated using (\ref{upar}) as
\begin{equation}
   L_{\pm} = \int\limits_{-\infty}^{\infty} Q_{\pm} {\cal F}(u_{\pm} + 
             Q_{\pm} x) d x    \quad .
\label{Lpm}
\end{equation}
The transformations above could also be used in the context of SPM, where
the integration in (\ref{Lpm}) is performed in an approximate way using an
expansion of $S(u)$ at $u=u_{\pm}$ (see next subsection for more details). 
However, the expression obtained for $L$ within the SPM is only approximate.  
Since our intention is to perform an exact calculation of $L$ (within a 
controllable precision), the integration 
in (\ref{Lpm}) is done numerically using Gaussian quadrature. Moreover, it 
is sufficient to calculate only $L_{+}$. Indeed, using
\begin{eqnarray}
   f(u^*)       &=& -f^*(u),\quad Q^*_{\pm} = Q_{\mp}, \\
   e^{i S(u^*)} &=& (-1)^N \exp[2 \xi i] \left\{ e^{i S(u)} \right\}^*
\end{eqnarray}
one obtains
\begin{equation}
   L_{-}  =  - (-1)^N \exp\left[2\xi i\right] L^*_{+}   \quad .
\label{Lm}
\end{equation}
Substituting (\ref{Lm}) into (\ref{Lsum}) yields
\begin{equation}
   L  =  L_{+}  + (-1)^N \exp\left[2 \xi i\right] L^*_{+}  \quad .
\label{Lexp}
\end{equation}
Introducing the absolute value ${\cal L}$ and the argument $\Omega$ of $L_{+}$
we obtain for the generalized Bessel function
\begin{equation}
   J_N(\xi, -z/2)  =  2 {\cal L} \cos(\xi - \Omega - N\pi/2)  
\label{JBes}
\end{equation}
and  
\begin{equation}
   |L|^2  =  2 {\cal L}^2 \left[ 1 + \cos(2 \xi - 2 \Omega - N\pi) \right].
\label{L2exp}
\end{equation}

We stress that no approximations have been done. The highly 
oscillatory integral (\ref{Lp}) has only been modified to a form that is 
much more convenient for the numerical calculation of $|L|^2$ and will be 
used in the present work to obtain numerical values of the SFA ionization 
rate $W_{\rm SFA}$ (\ref{W}) that serve as a reference for the QSFA 
derived below.

\subsection{\label{subsec:Bessel}Generalized Bessel functions in 
     the quasistatic limit}
%
In order to derive (in Sec.\,\ref{subsec:QSFA}) an analytic expression 
for the SFA rate in the quasistatic limit, it is required to first 
find the exact limit of $|L|^2$ for $\gamma \rightarrow 0$.
It follows from (\ref{upm}) that $u_{+}\rightarrow 0$ in the 
limit $\gamma\rightarrow 0$. Function $f(u)$ is then nearly 1
in the interval $(0,u_{+})$ and $S(u_{+})$ can be calculated  
using the Taylor expansion of $f^{-1}(u)$ at
$u=0$ for ${\rm Im}(u) >0$,
\begin{equation}
   f^{-1}(u) = 1 + \frac{u^2}{2} + \frac{3 u^4}{8} +\dots
\label{fexp}
\end{equation}
Substitution of (\ref{fexp}) into (\ref{Su}) and integration yields 
\begin{equation}
   i S(u_{+}) = - \frac{\tau \rho^3}{3} 
              - \frac{\tau \rho^3 \chi (3+\chi^2)}{6}\, i  + O(\gamma^2) \;.
\label{iSup}
\end{equation}
Performing the Taylor expansion of $S(u)$ at $u = u_{+}$ gives 
\begin{equation}
   i S(u_{+} +  Q_{+} x) = i S(u_{+}) - x^2 + i
                          \frac{\sqrt{2}}{3\sqrt{\tau\rho^3}} x^3 + O(\gamma^2)
\label{iSux}
\end{equation}
where
\begin{equation}
   Q_{+} = \gamma \sqrt{\frac{2 f(u_{+})}{\tau \rho }} = 
           \gamma \sqrt{\frac{2}{\tau \rho }} + O(\gamma^3)
\end{equation}
has been used. Due to the smallness of $Q_{+}$, function $e^{i S(u)}$ 
decays fastly in the vicinity of $u_{+}$. Since the expansions (\ref{iSux}) 
and (\ref{fexp}) are expected to be valid in this region, the integrand  
of $L_{+}$ can be rewritten as
\begin{equation}
   Q_{\pm} {\cal F}(u_{\pm} + Q_{\pm} x) = \gamma\, C\, 
           \exp\left[ -x^2 + i \frac{\sqrt{2} x^3}{3\sqrt{\tau\rho^3}} \right] 
           + O(\gamma^3)
\label{Integr}
\end{equation}
where
\begin{equation}
   C = \frac{1}{2\pi}\sqrt{\frac{2}{\tau \rho }} 
       \exp\left[ - \frac{\tau \rho^3}{3} -  
                  \frac{\tau \rho^3 \chi (3+\chi^2)}{6}\, i \right]  \;.
\end{equation}
Integration over $x$ yields for the 
absolute value ${\cal L}$ and the argument $\Omega$ of $L_{+}$
\begin{eqnarray}
\label{L_ex}
   {\cal L} &=& \gamma \frac{ \rho}{\sqrt{3}\pi} 
                K_{1/3}\left(\frac{\tau \rho^3 }{3}\right) + O(\gamma^3) \\
   \Omega   &=& -  \frac{\tau \rho^3 \chi (3+\chi^2)}{6}\,  + O(\gamma^2)
\label{Omega}
\end{eqnarray}
where $K_{\nu}$ is the modified Bessel function of the second kind of  
order $\nu$. Before using this result for a derivation of the ionization 
rate in the quasistatic limit, it is instructive to compare it to the 
predictions of the 'asymptotic approximation'~\cite{sfa:reis80} and the 
'tunneling 1 approximation'\cite{sfa:reis03} which both are based on the 
SPM.

It is important to stress that the term proportional to $x^3$ is usually 
ignored, if the SPM is used. Ignoring this term in 
(\ref{Integr}) and integrating over $x$ would yield instead of (\ref{L_ex}) 
\begin{eqnarray}
   {\cal L}_{\rm SPM} &=&   \frac{\gamma}{\sqrt{2\pi \tau \rho }} 
                            \exp\left[ - \frac{\tau \rho^3}{3}\right]. 
\label{L_SPM}
\end{eqnarray}
Although the limit $\gamma\rightarrow 0$ of the 'asymptotic approximation' 
is not easily transparent from the equations given in~\cite{sfa:reis80}, 
a tedious analysis yields exactly the form given in Eq.(\ref{JBes})  
where $\Omega$ is specified in Eq.(\ref{Omega}) and ${\cal L}$ must be 
substituted with ${\cal L}_{\rm SPM}$ from (\ref{L_SPM}). On the other 
hand, to obtain the limit $\gamma\rightarrow 0$ of the 'tunneling 1 
approximation' one should substitute ${\cal L}$ and $\Omega$ in 
Eq.(\ref{JBes}) with
\begin{eqnarray}
   {\cal L}_{\rm Tun1} &=& \frac{\gamma}{\sqrt{2\pi \tau \bar{\rho} }} 
                           \exp\left[ - \frac{\tau \bar{\rho}^3}{3}\right]  
                           + O(\gamma^3) \\
   \Omega_{\rm Tun1}   &=& -  \frac{\tau q_N \zeta \bar{\rho}^2}{2}  
                           + O(\gamma^2)
\label{OmegaTun1}
\end{eqnarray}
where $\bar{\rho} = \sqrt{1 + q_N^2}$.

\begin{figure}[!]
\includegraphics[width=80mm]{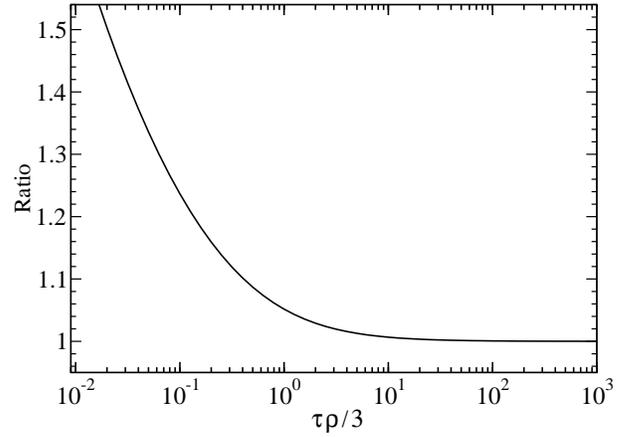}
\caption{\label{fig:Ratio}   
Ratio between ${\cal L}_{\rm SPM}$ (\ref{L_SPM}) and ${\cal L}$ (\ref{L_ex}) 
as a function of $\tau\rho^3/3$. This ratio indicates the range of validity 
of the SPM. Since $\rho \ge 1$, the SPM result shows very good agreement 
for large $\tau$ and starts to fail only for small values of $\tau$. The  
latter case corresponds to a small binding energy or a high intensity of the 
field.
} 
\end{figure}

Clearly, the 'tunneling 1 approximation' agrees with the 'asymptotic 
approximation' only for $\zeta \approx 0$. It is also evident that for 
$\zeta \approx 1$ the damping factor in the exponent increases with $q_N$, 
whereas for the 'asymptotic approximation' it remains at about $-\tau/3$.
Since $\zeta \approx 1$ gives the main contribution to ionization, 
the 'tunneling 1 approximation' underestimates the ionization 
rate for larger kinetic momenta as is numerically proven 
in~\cite{sfa:baue05c}.

It is also instructive to check the range of the validity of the SPM used 
to obtain the 'asymptotic approximation'. The ratio between 
${\cal L}_{\rm SPM}$ and ${\cal L}$ is given by
\begin{equation}
   \frac{{\cal L}_{\rm SPM}}{{\cal L}} =   
            \frac{\sqrt{\pi}\exp[- \tau \rho^3/3]}{
                \sqrt{2/3}\sqrt{\tau \rho^3} K_{1/3}(\tau \rho^3 /3)}.
\end{equation}
Its numerical values for different parameters is shown  
in Fig.~\ref{fig:Ratio}. For large values of $\tau\rho^3/3$ the ratio 
is given asymptotically as 
\begin{equation}
   \frac{{\cal L}_{\rm SPM}}{{\cal L}} \rightarrow  
                          1 + \frac{5}{72}\frac{3}{\tau \rho^3} 
                          =  1 + \frac{5 F}{24\kappa^3 \rho^3}
\end{equation}
and approaches 1 in the weak-field limit. For usual laser parameters SPM may 
give an error within a few percent and only in the extreme case of small 
binding energies (e.\,g.\ ionization of Rydberg states) and very strong 
fields the error is significantly larger.

\subsection{\label{subsec:QSFA} SFA rate in the quasistatic limit}

Having obtained an exact asymptotic expression for $|L|^2$ it is now possible 
to derive an analytic form of the SFA ionization rate $W_{\rm SFA}$ 
(\ref{W}) in the quasistatic limit. Besides the formulas obtained in the 
previous two subsections some further asymptotically exact approximations  
are, however, required. For this purpose, defining the azimuthal angle $\phi$
around axis parallel to $\vek{F}$ and using 
\begin{equation}
  \int  d \hat{\vek{p}} = \int\limits_{-1}^{1}  d \zeta 
                          \int\limits_{0}^{2\pi}  d \phi  
\end{equation}
we rewrite Eq.(\ref{W})  as
\begin{eqnarray}
   W_{\rm SFA} &=& \frac{\kappa^5}{16\pi^2} \int\limits_{-1}^{1}  d \zeta  \,
                   \sum_{N \ge N_{0}}\,  q_N  \left( 1 + q_N^2  \right)^2 
                   | L |^2\,\tilde{\Phi} 
\label{Wr}
\end{eqnarray}
where
\begin{equation}
   \tilde{\Phi} = \int\limits_{0}^{2\pi}  d \phi  |\tilde{\Psi}_0|^2.
\end{equation}

Note, the argument of the cosine in Eqs.(\ref{L2exp}) is proportional
to $\gamma^{-3}$ and leads to fast oscillations, if $N$ and $\zeta$ are 
varied. Thus the contribution of this term to the final result is negligibly 
small and it is possible to substitute $| L |^2$ in (\ref{Wr}) with 
$2{\cal L}^2$, 
\begin{eqnarray}
   W^{\rm ap 1}_{\rm SFA} &=& \frac{\kappa^5}{8\pi^2} 
                 \int\limits_{-1}^{1}  d \zeta  \, \sum_{N \ge N_{0}}\,  q_N  
                 \left( 1 + q_N^2  \right)^2 {\cal L}^2\tilde{\Phi} \;.
\label{ap1SFA}
\end{eqnarray}

The next step is to substitute the summation over $N$ by an integral. A 
standard approach consists of a transformation of the sum into an integral 
over $q_N$. This allows to calculate differential rates, but due to the 
coupling of $q_N$ and $\zeta$ in $\rho$ it is impossible to obtain a 
simple analytical expression without the use of an expansion (e.\,g.\ the 
small kinetic momentum one, $q_N \ll 1$). Instead of the use of a 
double integral with respect to $q_N$ and $\zeta$ we rewrite (\ref{ap1SFA}) 
as a double integral with respect to $\rho$ and $\chi$. Transforming the 
sum into the integral over $\rho$ with
\begin{equation}
   \sum_{N\ge N_0}  \rightarrow \int\limits_{1}^{\infty} d \rho 
                                \frac{ \rho\tau}{\gamma (1 - \zeta^2)},
\end{equation}
and using
\begin{equation}
   1 + q_N^2 = \rho^2 (1 + \chi^2),\quad
  \int\limits_{-1}^{1}  d \zeta\, \frac{ q_N }{  (1 - \zeta^2)} = 
                                   \int\limits_{-\infty}^{\infty}  d \chi\,\rho
\end{equation}
one obtains
\begin{equation}
   W^{\rm ap 2}_{\rm SFA} = \frac{\kappa^5\tau}{8\pi^2\gamma} 
                            \int\limits_{1}^{\infty} d \rho\, \rho^6  
                            \int\limits_{-\infty}^{\infty}  d \chi
                            (1+ \chi^2)^2 {\cal L}^2(\rho,\chi) 
                            \tilde{\Phi}(\rho,\chi) \;.
\label{ap2SFA}
\end{equation}
Substitution of (\ref{L_ex}) into (\ref{ap2SFA}) yields an  
analytical expression for the quasistatic limit of the SFA (denoted QSFA) 
\begin{equation}
\label{W_QSFA}
   W_{\rm QSFA}   = \omega \,  R(\tau, \kappa)
\end{equation}
with
\begin{eqnarray}
\label{R}
   R(\tau,\kappa) &=& \int\limits_{1}^{\infty} \frac{8 \tau^2}{3\pi}
                     B(\rho,\kappa) K^2_{1/3}\left(\frac{\tau \rho^3}{3}\right)
                     \,  d \rho \\
\label{Bcoef}
   B(\rho,\kappa) &=& \rho^8 \left(\frac{\kappa}{4\pi}\right)^3 
                     \int\limits_{-\infty}^{\infty}  d \chi\, (1+ \chi^2)^2\, 
                     \tilde{\Phi}(\rho,\chi) \;.
\end{eqnarray}

If the SPM is used, one has to use (\ref{L_SPM}) instead of 
(\ref{L_ex}) in (\ref{ap2SFA}). As a consequence, one obtains a  
similar result, but function $R(\tau,\kappa)$ has to be substituted with 
\begin{equation}
  R_{\rm SPM}(\tau,\kappa) = \int\limits_{1}^{\infty} \frac{4 \tau}{\rho^3}   
         B(\rho,\kappa) \exp\left[ -\frac{2\tau}{3}\rho^3\right]\, d \rho \;.
\label{RSFA}
\end{equation}

Eq.\,(\ref{W_QSFA}) is one of the central results of the present work. It 
shows that the ionization rate calculated within SFA is proportional to the 
frequency $\omega$ in the limit $\omega \rightarrow 0$. Thus the SFA rate 
vanishes in the static limit $\omega=0$ for all binding energies and field 
strengths! Clearly, this prediction of SFA is unphysical implying that 
SFA is not applicable in the quasistatic limit for atomic systems. As is 
also clear from the derivation, this conclusion is not a consequence of the 
usually adopted SPM or small-momentum approximation, since they were not 
adopted. 

The present finding appears to be in conflict with the discussion given, 
e.\,g., in \cite{sfa:beck05}, where it is explicitly stressed that  
SFA (corresponding to the first-order S-matrix theory in velocity gauge) 
approaches the correct tunneling limit for $\omega\rightarrow 0$ (and 
sufficiently weak fields). However, in \cite{sfa:beck05} (and corresponding 
references therein) this conclusion is reached on the basis of a derivation 
valid for short-range potentials, while the present result is obtained 
for the long ranged Coulomb potential. For short-range potentials the 
integral over $\chi$ in (\ref{Bcoef}) diverges and one has to consider the  
term proportional to $\gamma^2$ in (\ref{iSup}). This term 
removes the divergence and the obtained limit for the ionization rate is now 
$\omega$ independent in accordance with the discussion in \cite{sfa:beck05}.
For long-range potentials there is no divergence in (\ref{Bcoef}), and 
thus (\ref{W_QSFA}) gives the corresponding quasistatic limit of SFA 
for that case. Clearly, the agreement of the SFA rate with 
the one predicted by tunneling theories that is obtained for short-range 
potentials cannot be used as a measure of the validity of the SFA,  
if long-range potentials are present. However, as is discussed below 
for the specific example of hydrogen-like atoms, the QSFA results may be 
used together with tunneling theories to obtain an approximate Coulomb 
correction factor for SFA.   
  
Since for long-range potentials the SFA rate leads to unphysical results 
in the quasistatic limit, one would expect that there is very limited 
interest in its explicit calculation. However, as is shown below, the 
explicit calculation of $W_{\rm QSFA}$ is not only useful for obtaining 
a Coulomb correction factor, but it provides also an alternative recipe 
for an efficient though approximate calculation of SFA rates for atomic 
and molecular systems exposed to intense laser fields in a large range 
of experimentally relevant laser parameters. To demonstrate this, 
calculations for hydrogen-like atoms using Eqs.\,(\ref{W_QSFA}) to 
(\ref{Bcoef}) are discussed in the next section.

\section{\label{sec:hydrogen}Quasistatic limit of SFA for 
         hydrogen-like atoms}

\begin{figure}[!]
\includegraphics[width=80mm]{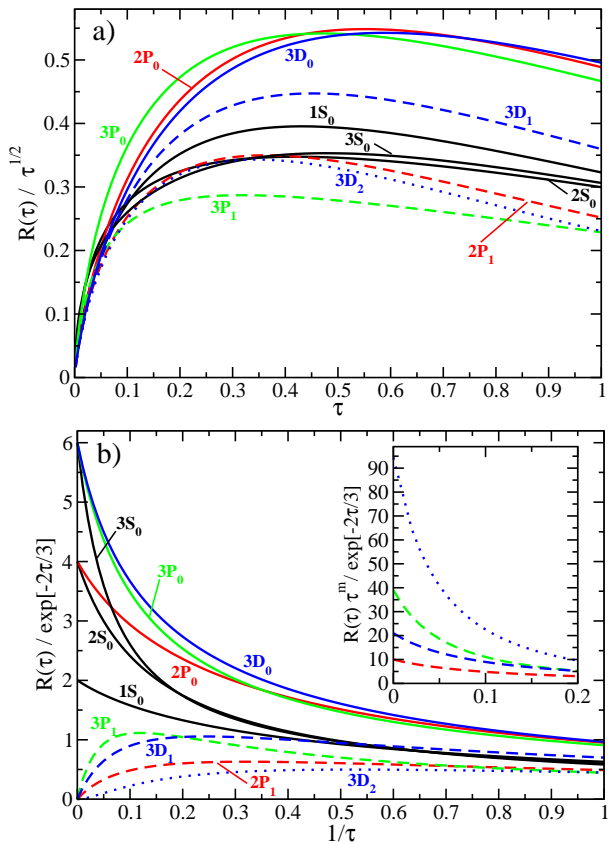}
\caption{\label{fig:R_for_H}  
(Color online) 
The proportionality coefficient $R(\tau) =  W_{\rm QSFA}/\omega$ (\ref{R})
is shown for the complete range of $\tau = (2 E_{\rm b})^{3/2}/F$ values 
for the different hydrogen-like states with $n\leq 3$.
a) For small $\tau$  (corresponds to a strong field $F$ or a small binding 
   energy $E_{\rm b}$) all coefficients decrease with decreasing $\tau$. 
   To partly compensate this effect all 
   coefficients $R(\tau)$ are scaled by factor $\sqrt{\tau}$.
b) For large $\tau$ (corresponds to a weak field $F$ or a large binding 
   energy $E_{\rm b}$) all coefficients are scaled by the 
   factor $\exp[-2\tau/3]$ (in the insert the coefficients for $m>0$ 
   are also shown scaled by the factor $\tau^{-m}\exp[-2\tau/3]$). 
   In this limit the coefficients $R(\tau)$ tend to 
   those given by (\ref{LimR}).
}
\end{figure}

\begin{figure*}[!]
\includegraphics[width=160mm]{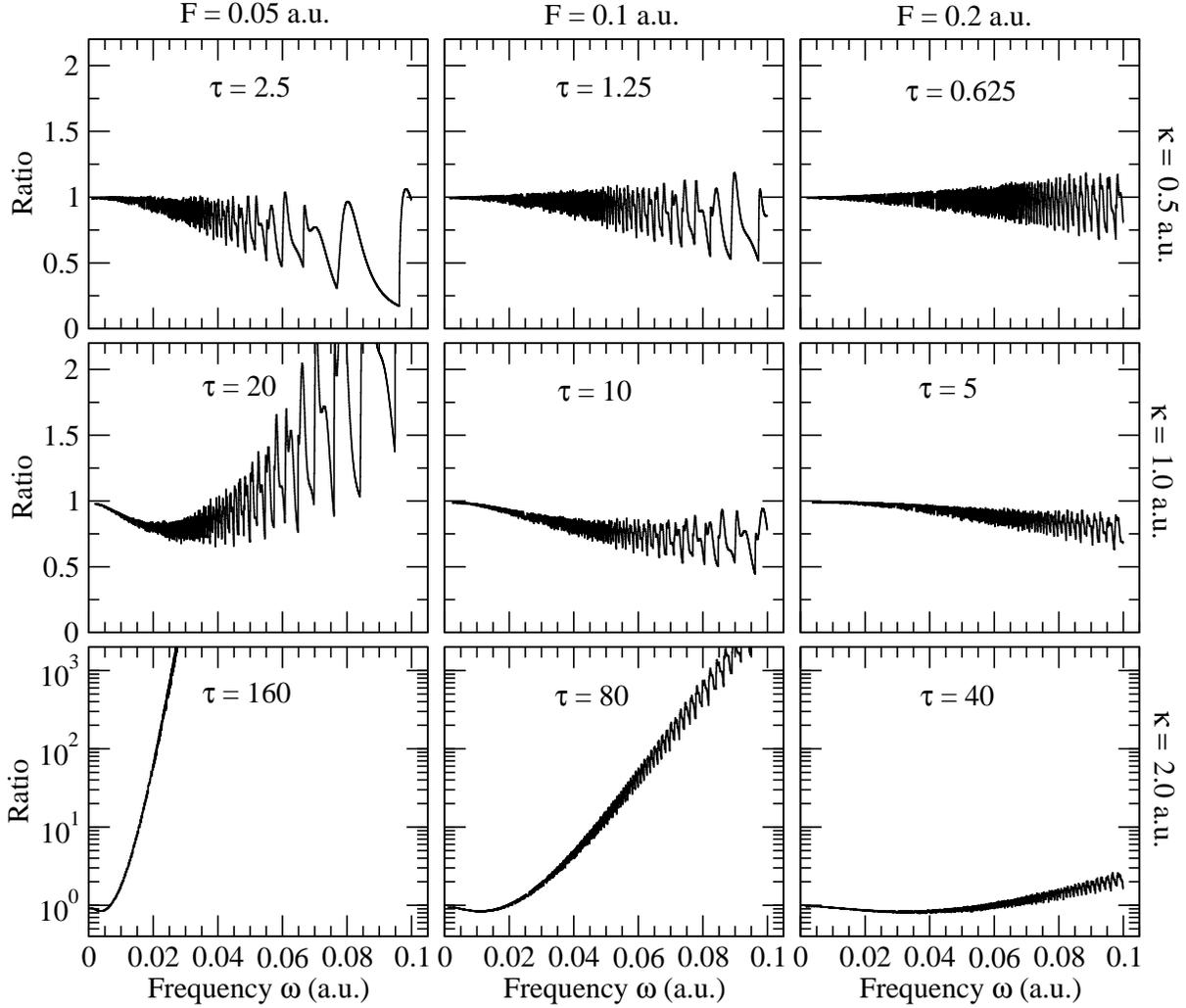}
\caption{\label{fig:Ratio_1S}  
Ratio $W_{\rm SFA}/W_{\rm QSFA}$ for the 1S state of a hydrogenlike atom as a 
function of the frequency $\omega$ for different field strength $F$ and  
electron binding energies $E_{\rm b} = \kappa^2/2$.
} 
\end{figure*}

\begin{figure*}[!]
\includegraphics[width=160mm]{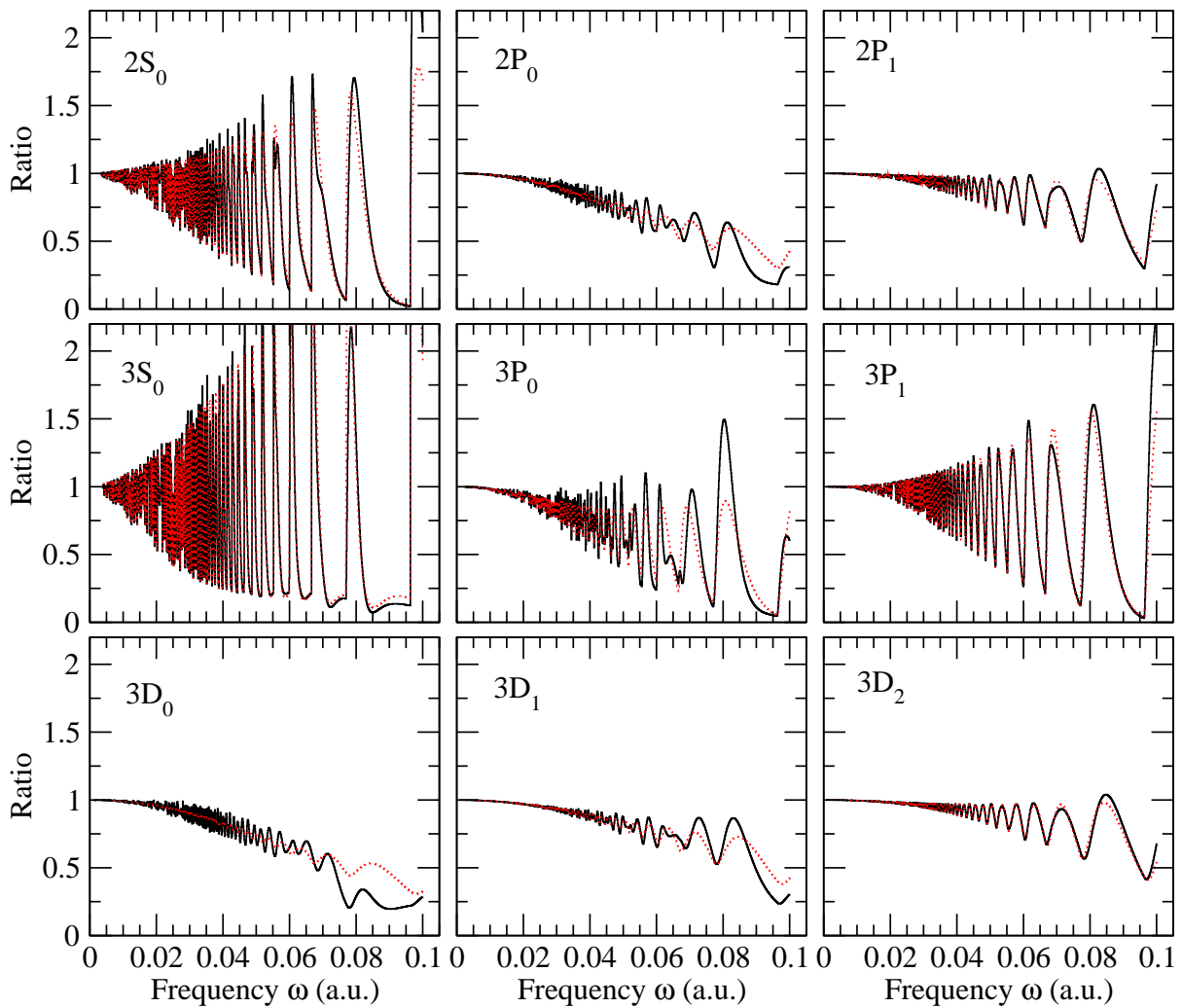}
\caption{\label{fig:Ratio_nS_1}  
(Color online)  
The ratios $W_{\rm SFA}/W_{\rm QSFA}$ (black solid) and
$W^{\rm ap1}_{\rm SFA}/W_{\rm QSFA}$ (red dotted) are shown as a function 
of the frequency $\omega$ for all atomic states ($n\leq 3$) and an electron
binding parameter $\kappa=0.5\,$a.\,u.\ at the field strength
$F=0.05\,$a.\,u. Since both curves are in relatively good agreement with 
each other, the 1st step in the derivation of the QSFA discussed in 
Sec.\ref{subsec:QSFA} is valid also for finite values of $\omega$.
} 
\end{figure*}

\begin{figure*}[!]
\includegraphics[width=160mm]{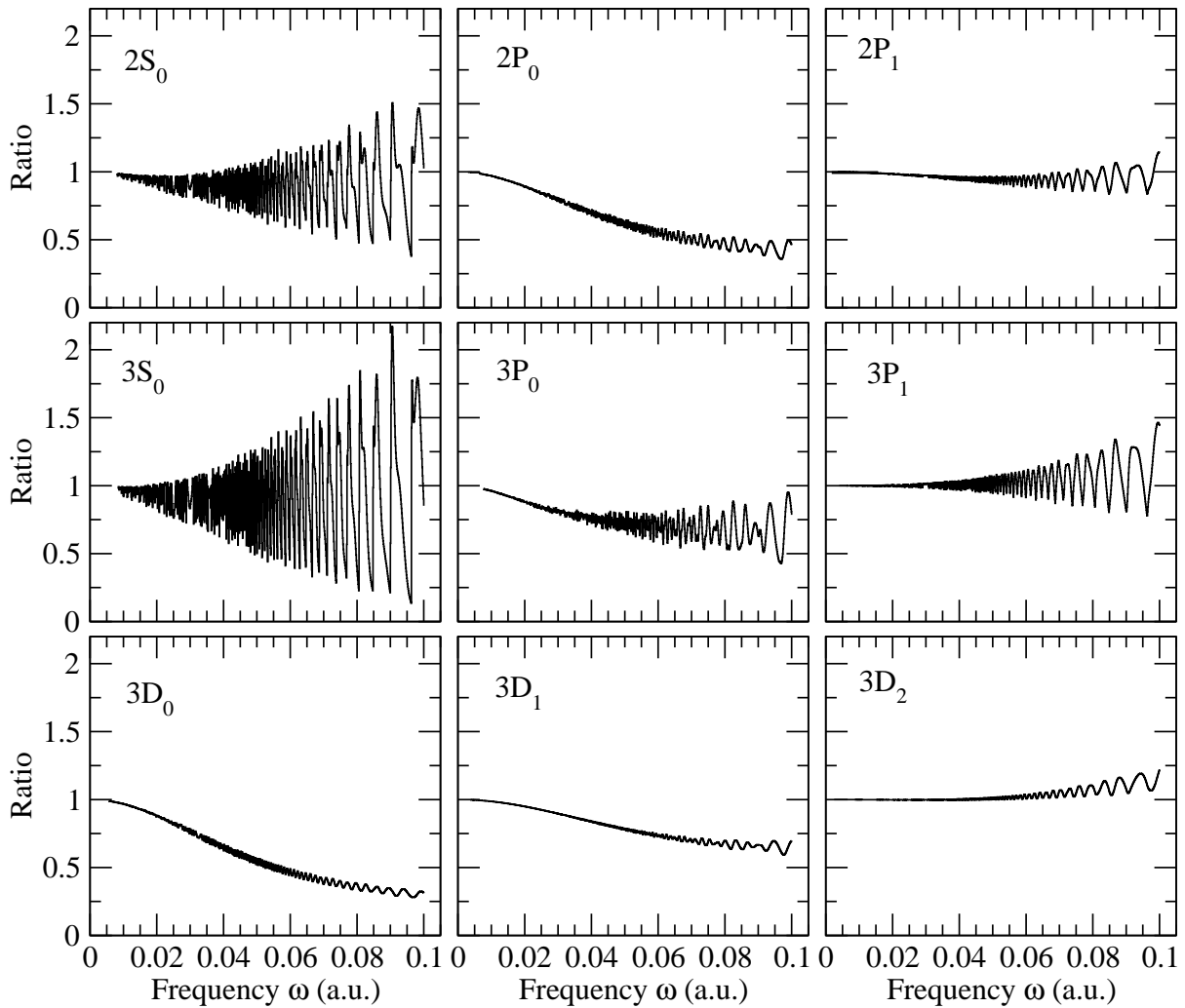}
\caption{\label{fig:Ratio_nS_2}  
As Fig.\,\ref{fig:Ratio_nS_2}, but for field strength $F=0.1\,$a.u. and 
the electron binding parameter $\kappa=1\,$a.\,u. (Only ratio 
$W_{\rm SFA}/W_{\rm QSFA}$ is shown.)
}
\end{figure*}

\begin{figure}[!]
\includegraphics[width=80mm]{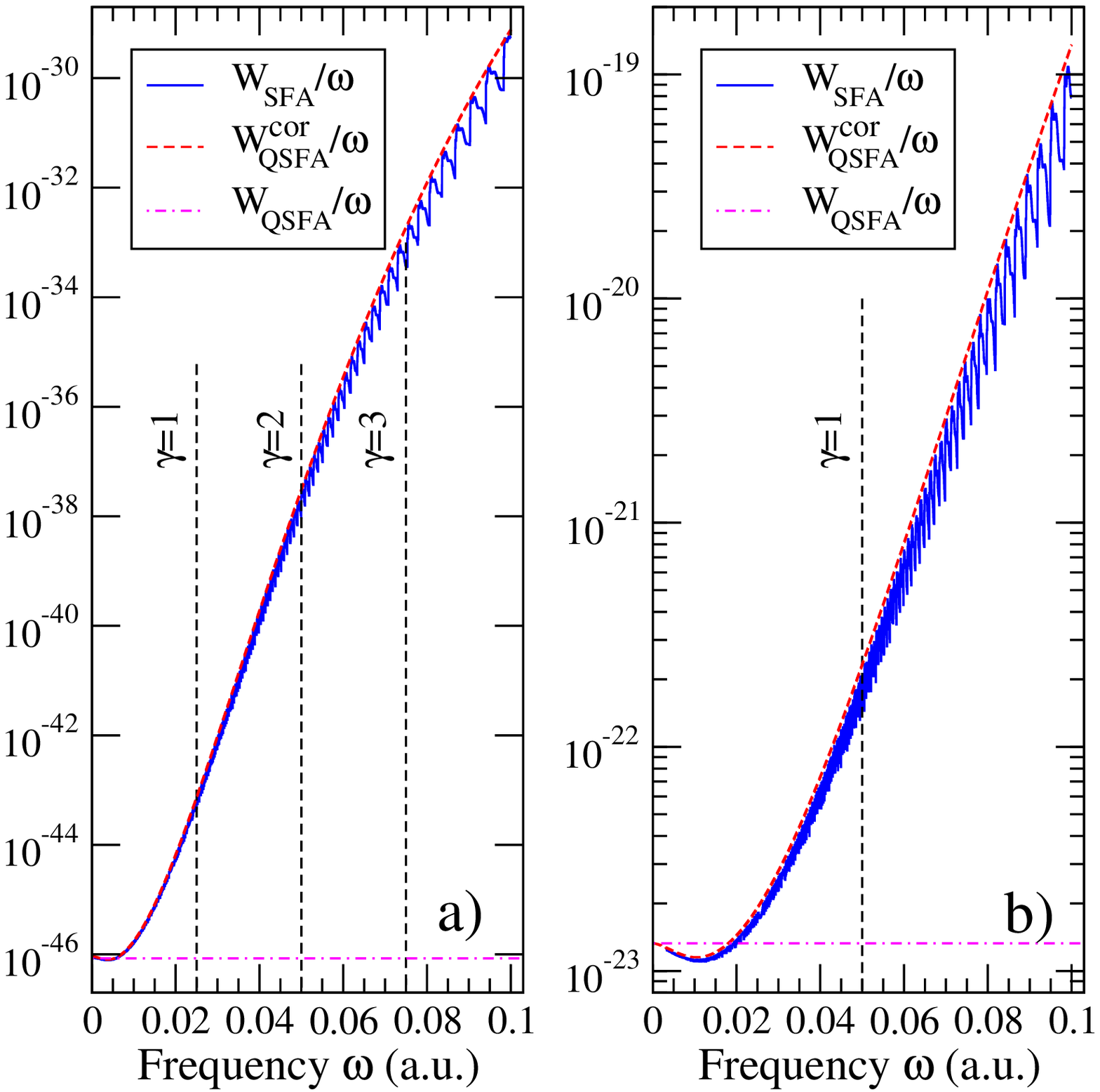}
\caption{\label{fig:Cor_QSFA}
(Color online)   
The exact SFA rate $W_{\rm SFA}$, the QSFA rate $W_{\rm QSFA}$, and
the corrected QSFA rate 
$W^{\rm cor}_{\rm QSFA}$ (all scaled by $\omega^{-1}$) are shown 
for the 1S state of a hydrogenlike atom with $\kappa = 2\,$a.\,u.\ at  
a) $F = 0.05\,$a.\,u.\ and b) $F = 0.1\,$a.\,u.  
The frequencies $\omega$ corresponding to various values of Keldysh  
parameter $\gamma$ are indicated with dashed lines. 
} 
\end{figure}

\subsection{\label{subsec:Coeff}Proportionality coefficient $R$}

In the case of bound states of hydrogen-like atoms the integral over 
$\chi$ in (\ref{Bcoef}) can be analytically calculated using the identity
\begin{equation}
   \int\limits_{-\infty}^{\infty}  \frac{d \chi}{ (1+ \chi^2)^n} = 
                                                \pi \frac{ (2n-3)!!}{(2n-2)!!}.
\end{equation}  

For example, the Fourier transform of the 1S$_0$ state is given by
\begin{equation}
   \tilde{\Psi}_0 = \frac{16 \pi}{\kappa^{3/2}}\frac{1}{(1 + q^2_N)^2} 
                    \frac{1}{\sqrt{4\pi}}
\label{FT_H}
\end{equation}
resulting in 
\begin{equation}
   \tilde{\Phi}(\rho,\chi) =  \frac{2^7 \pi^2}{\kappa^3 
                                      \rho^8 (1 + \chi^2)^4} \; .
\label{PhiH}
\end{equation}
Substitution of (\ref{PhiH}) into (\ref{Bcoef}) and integration over $\chi$ 
yields $B_{\rm 1S_0} = 1$. The functions $B(\rho)$ for all hydrogenic states 
with principal quantum number $n\leq 3$ are listed in Appendix~\ref{app:Bcoef}.
Note, for hydrogen-like states the function $B(\rho)$ is 
$\kappa$-independent and, therefore, the proportionality coefficient $R$ is
a function of $\tau$ only. The evaluation of $R$ according to Eq.\,(\ref{R}) 
can then simply be performed numerically. Since the integrand is a smooth   
exponentially decaying function [this is directly evident, if the SPM 
approximation is adopted as in (\ref{RSFA})], quadrature can easily 
and very efficiently be performed with high precision. 

The proportionality coefficients $R(\tau)$ for a variety of states of
hydrogen-like atoms are shown in Fig.~\ref{fig:R_for_H} for the  
complete range of values of the inverse field parameter $\tau$. 
In Fig.~\ref{fig:R_for_H}\,a the range $\tau\leq 1$ is shown. For 
better visibility 
the function $R(\tau)/\sqrt{\tau}$ is plotted instead of $R(\tau)$.      
It is worth noticing that for very small $\tau$ the $R$ values for 
all states approach 0. This is a known failure of SFA, since a larger 
field intensity to binding energy ratio (and thus smaller $\tau$) should 
clearly result in a larger and not in a smaller ionization 
rate~\cite{sfa:baue06}. %

In the $\tau$ range shown in Fig.~\ref{fig:R_for_H}\,a in which the $R$ 
values follow the expected behavior (decreasing $R$ for increasing 
$\tau$), the different states vary rather differently as a function of 
$\tau$. It is clearly visible that $R$ depends mostly on the quantum 
numbers $l$ and $m$ and only very weakly on $n$. A different dependence  
is found for large values of $\tau$ as is discussed below.

\subsection{\label{subsec:weak}Weak-field limit}
%
The weak-field limit $F\rightarrow 0$ corresponds to $\tau \rightarrow 
\infty$. Using an asymptotic expansion for the modified Bessel function one 
finds that the integrand in (\ref{R}) is proportional 
to $\exp[-2\tau\rho^3/3]$. For large values of $\tau$ the integrand 
decays thus rapidly as $\rho$ increases. Therefore, it is possible to use 
an expansion in terms of $\rho$ at $\rho=1$. This procedure yields 
\begin{equation}
   R_{nlm}(\tau) \rightarrow C^{\rm QSFA}_{nlm} (2\tau)^{-|m|} e^{-2\tau/3 }.
\label{LimR}
\end{equation}
The general expression for the coefficients $C^{\rm QSFA}_{nlm}$ is quite 
complicated. A very simple result occurs, however, for $m=0$ where 
$C^{\rm QSFA}_{nl0}=2n$ is obtained. Note, in this case the coefficient is 
$l$-independent. For $n\le 3$ the coefficients are given by
\begin{equation}
   C^{\rm QSFA}_{nlm}= 2n \frac{[2n(n-l)+2|m| - 1]!!}{[2n(n-l) - 1]!!} \;.
\end{equation}
Noteworthy, the $l$ dependence is in fact limited to the circularity 
of the hydrogenic state, since $l$ appears only in the form $n-l$. 

In Fig.\ref{fig:R_for_H}\,b function $R$ is shown (after multiplying it 
with $e^{2\tau/3 }$ to remove the exponential dependence on $\tau$) as 
a function of $1/\tau$. The weak-field limit corresponds thus to 
$1/\tau \rightarrow 0$. As predicted, for $m=0$ the scaled function $R$ 
approaches $C^{\rm QSFA}_{nl0}=2n$ in this case. Due to the 
$\tau^{|m|}$ factor appearing in (\ref{LimR}) the high $|m|$ states 
are harder to ionize in the weak-field limit. It is also apparent  
from Fig.\ref{fig:R_for_H}\,b that the characteristic dependence on the 
quantum numbers in the weak-field limit is reached only for $1/\tau \leq 0.1$ 
to 0.2. For example, down to $1/\tau \approx 0.15$ the QSFA ionization rates 
of the 2S$_0$ and 2P$_0$ states are almost identical.

It is instructive to compare the quasistatic limit of SFA ionization rate in 
weak-field limit with the well-known quasistatic 
Popov-Peremolov-Terent'ev (PPT) formula~\cite{sfa:pere66}, 
\begin{equation}
   W_{\rm PPT} = |C_{nl}|^2\, f_{lm}\, \sqrt{\frac{3}{2\pi}}\,\kappa^2\,
                 (2\tau)^{2n-|m|-3/2}\, e^{-2\tau/3 }
\label{WPPT}
\end{equation}
where
\begin{equation*}
   |C_{nl}|^2\, = \frac{2^{2n}}{n(n+l)! (n-l-1)!},\quad f_{lm} 
                = \frac{(2l+1)(l+|m|)!}{2^{|m|} |m|! (l-|m|)!} \;.
\end{equation*}

The ionization rates $W_{\rm QSFA}$ and $W_{\rm PPT}$ both include the 
exponential term $\exp[-2\tau/3]$ and the factor $(2\tau)^{-|m|}$, 
but differ in the remaining part. Introducing the ratio
\begin{equation}
   Q_{\rm PPT} = \frac{W_{\rm PPT}}{W_{\rm QSFA}} 
               = \frac{1}{\omega}\frac{2^{2n-2} |C_{nl}|^2\, 
                 f_{lm}}{C^{\rm QSFA}_{nlm}}\sqrt{\frac{3}{\pi}} 
                 \frac{F^{3/2}}{\kappa^{5/2}} \frac{\kappa^{6n}}{F^{2n}}
\label{Rcc}
\end{equation}
it is possible to identify four factors that prevent an agreement 
between the QSFA and PPT predictions. One is due to the (unphysical) 
$\omega$ dependence of QSFA. Also the constant factors that depend 
on the quantum numbers $n$, $l$, and $m$ differ. For example, for 
fixed $n$ and $m=0$ QSFA predicts the same ionization rate for 
states with different $l$ whereas the PPT formula predicts an $l$ 
dependence. Then there is a constant factor ($\sqrt{3/\pi}$) that is, 
however, very close to 1. Finally, both rates differ in their dependence 
on field strength and binding energy which is expressed as two factors 
to stress the $n$ dependence or independence. 

Since QSFA is the exact asymptotic limit of SFA, Eq.\,(\ref{Rcc}) 
can be used to derive a Coulomb-corrected SFA rate,   
$W_{\rm CSFA} = Q_{\rm PPT} \, W_{\rm SFA}$. Clearly, the factor 
$Q_{\rm PPT}$ derived here explicitly for atomic hydrogen could be 
applied also to other atomic or molecular systems by performing the 
evident modifications like the introduction of effective quantum 
numbers~\cite{sfa:ammo86}, such as $n^*$, $l^*$, etc. %
Although the range of validity of $W_{\rm CSFA}$ for 
$\omega\neq 0$ is not directly evident, in contrast to $W_{\rm SFA}$ it 
at least reaches the tunneling limit. Already in the past efforts have 
been made to derive Coulomb-correction factors for KFR theories, but so 
far the resulting rates did not lead to convincing results 
(see~\cite{sfa:beck05} and references therein). Based on some approximations, 
A. Becker {\it at al.}~\cite{sfa:beck01} have proposed a Coulomb correction 
factor, $C^2$. Using this factor, very good agreement is found between 
experimental and theoretical SFA ionization yields for a large number of 
atoms and laser frequencies. The comparison is, however, mostly performed 
on a qualitative level, since the experiments did not provide absolute 
yields and thus the theoretical and experimental data were adjusted at 
one common point. In addition, SFA results for atomic hydrogen (with and 
without $C^2$ factor) are compared to full numerical solutions of the 
time-dependent Schr\"odinger equation in~\cite{sfa:beck01} and again 
good agreement is found (on logarithmic scale).  

For atomic hydrogen one has $ C^2 = \kappa^{6n}/F^{2n}$ which corresponds 
just to the last factor in (\ref{Rcc}). Clearly, the $C^2$-corrected 
SFA rate does not approach the tunneling limit for $\omega \rightarrow 0$.  
However, the terms missing in $C^2$ yield for 
$\omega = 0.05\,$a.\,u.\ and the ground state of a hydrogen atom a 
factor 0.5 - 1.2 for $F = 0.05 - 0.1\,$a.\,u. This can explain the 
reasonable agreement of the $C^2$-corrected SFA results with the 
ones of {\it ab initio} calculations reported for such parameters 
in~\cite{sfa:beck01}. However, the deviation increases by a factor 10 for 
the CO$_2$ laser frequency or for larger $n$. It may be noted that 
although~\cite{sfa:beck01} contains also comparisons with experimental 
data obtained with a CO$_2$ laser, the present work shows that the found 
agreement is due to the fact that the comparison is made on a relative 
scale, as mentioned before. In this case the erroneous $\omega$ dependence 
of SFA (clearly not corrected by the $C^2$ factor) is, for example, not 
visible.

\subsection{\label{subsec:validity}Range of validity of QSFA}

As follows from the derivation, $W_{\rm QSFA}$ in (\ref{W_QSFA}) is the 
exact asymptotic form of the SFA ionization rate $W_{\rm SFA}$ in the limit 
$\omega \rightarrow 0$. It is of course interesting to investigate the 
validity regime of QSFA for non-zero values of $\omega$. In fact, as is 
shown now, QSFA provides for a wide range of parameters a good approximation 
to SFA even for laser wavelengths of around 800\,nm or less.  

Fig.~\ref{fig:Ratio_1S} shows the ratio $W_{\rm SFA} / W_{\rm QSFA}$ 
for the 1S state of a hydrogenlike atom as a function of laser frequency 
for nine different values of the inverse field parameter $\tau$. The 
variation of $\tau$ is achieved by using three different values for both the   
binding-energy related quantity $\kappa$ and the field intensity $F$. 
The (reference) ionization rate $W_{\rm SFA}$ has been calculated numerically 
using the scheme described in Sec.\,\ref{subsec:Efc}. All curves approach 
unity for $\omega \rightarrow 0$ indicating the correctness of the 
derivation of QSFA as well as numerical consistency. In the case of the 
smallest shown value of the inverse field parameter, $\tau =0.625$, one 
notices that the ratio shows an oscillatory behaviour that is due to channel 
closings that are not resolved in QSFA. The oscillation amplitude 
increases with $\omega$, but the ratio remains in between about 0.75 and 
1.25 in the full frequency range. Therefore, QSFA is correct to within 
25\%. If one averages over the oscillations, one finds an even much 
better quantitative agreement between QSFA and SFA. In view of the fact 
that the SFA rate is known to overestimate the effect of channel closings 
and that these pronounced channel closing features mostly disappear when 
averaging over realistic laser parameters (envelope, focal volume etc.), 
the QSFA can be said to provide a very accurate approximation for the 
given parameters. Note, $\omega = 0.1\,$a.\,u.\ corresponds to a laser 
wave length of about 450\,nm and thus the shown frequency range covers 
a large range of experimentally relevant lasers. 

Increasing $\tau$ by decreasing $F$ (but keeping $\kappa$ fixed) leads 
to larger oscillation amplitudes while the oscillation period increases. 
Most importantly, the average value of the ratio drops with increasing 
$\omega$ below 1. QSFA starts to overestimate the SFA rate. Nevertheless, 
the oscillation averaged QSFA rate deviates for $\tau=2.5$ at 800\,nm 
from SFA by less than 25\,\%. Increasing $\tau$ by increasing $\kappa$ 
(for fixed $F=0.2\,$a.\,u.) decreases the oscillation amplitude. However, 
the $\omega$ averaged ratio deviates more from unity than for smaller 
binding energies. Changing $\kappa$ from 0.5\,a.\,u.\ (corresponding to 
a binding energy $E_b\approx 3.5\,$eV) to 1.0\,a.\,u.\ (13.6\,eV, 
hydrogen atom) and 2.0\,a.\,u.\ (27.2\,eV, He$^+$) changes the ratio  
at $\omega \approx 0.1\,$a.\,u.\ to about 0.75 and 2.0, respectively.  
From the representative examples shown in Fig.~\ref{fig:Ratio_1S} 
one can see that these are general trends. A decrease of $F$ 
(fixed $\kappa$) leads to larger oscillation amplitudes and deviations 
of the averaged ratio from unity. This limits the applicability of QSFA 
to a smaller $\omega$ range. An increase of the binding energy (fixed 
$F$) damps the oscillation, but increases the deviation from unity. 
Combining both results it is clear that QSFA works best for small 
binding energies and high field strengths and thus for small values 
of $\tau$. However, $\tau$ alone is not a sufficient parameter to 
describe the validity of QSFA, as can be seen from the examples 
shown for $\tau=2.5$ and 5.0. In this example QSFA works better 
for the larger value of $\tau$ that is realized by enlarging 
both $E_b$ and $F$.

In Figs.\,\ref{fig:Ratio_nS_1} and \ref{fig:Ratio_nS_2} the 
validity of the QSFA is investigated for different initial states 
of hydrogen-like atoms. This includes all possible states with $n\leq 3$. 
Fig.\,\ref{fig:Ratio_nS_1} shows the results for $F=0.05\,$a.\,u.\ and 
$\kappa=0.5\,$a.\,u. These are the same parameters as the ones used for 
the 1S state in the upper left corner of Fig.\,\ref{fig:Ratio_1S}. 
The results in  Fig.\,\ref{fig:Ratio_nS_2} were on the other hand 
obtained with $F=0.1\,$a.\,u.\ and $\kappa=1.0\,$a.\,u.\ and 
correspond therefore to the ones in the middle of Fig.\,\ref{fig:Ratio_1S}. 
Again, all ratios approach unity for $\omega \rightarrow 0$ as it should 
be. Comparing the results for the S states one notices that the oscillation 
amplitude increases with $n$, but the deviation of the $\omega$-averaged 
results is very similar. The same trend is visible within the P states 
(for either $m=0$ or $m=1$). For a given $n$ value the oscillations 
are most pronounced for $l=0$ and decrease with increasing $l$. In view 
of the $\omega$-averaged results the range of validity of the QSFA as a 
function of $\omega$ shows, however, a weaker dependence on $l$, but is 
in fact decreasing for increasing $l$. For a given $n$ and $l$ combination 
(2P$_0$ and 2P$_1$, 3P$_0$ and 3P$_1$, or 3D$_0$, 3D$_1$, and 3D$_2$) 
the $\omega$-averaged ratios indicate that the range of validity of 
the QSFA increases with $m$. One may notice the close similarity of the 
results within the series 1S$_0$, 2P$_1$, and 3D$_2$, 2P$_0$ and 3D$_1$, 
as well as 2S$_0$ and 3P$_1$. Finally, as was the case for the 1S state, 
also Figs.\,\ref{fig:Ratio_nS_1} and \ref{fig:Ratio_nS_2} show that a larger 
value of $\tau$ decreases the oscillation amplitude and increases the 
validity regime of the QSFA.      

Fig.~\ref{fig:Ratio_nS_1} shows in addition the ratio of 
$W^{\rm ap1}_{\rm SFA}$ [see Eq.(\ref{ap1SFA})] and $W_{\rm QSFA}$. 
The overall good agreement with the ratio $W_{\rm SFA}/W_{\rm QSFA}$ 
indicates $W^{\rm ap1}_{\rm SFA} \approx W_{\rm SFA}$. Clearly, the first 
step in deriving QSFA is well justified for finite frequencies. 
Especially, the highly oscillatory behavior of the rate due to channel 
closings is relatively well reproduced by $W^{\rm ap1}_{\rm SFA}$. 
The main reason for the deviation between SFA and QSFA is thus due to 
step 2 of the derivation which smoothes out the highly oscillatory 
behavior of the rate if $\omega$ is varied.

It is instructive to investigate the main reason for the failure of QSFA 
to reproduce SFA for large values of $\tau$. It turns out that for large 
$\tau$ it is most essential to consider in Eq.\,(\ref{iSup}) also the 
terms proportional to $\gamma^2$. This yields (see Appendix~\ref{app:Cor}) 
\begin{equation}
   W^{\rm cor}_{\rm QSFA} = C_{\rm cor}  W_{\rm QSFA}
\label{cQSFA}
\end{equation}
where the correction factor (for a hydrogenlike 1S state) is given by
\begin{equation}
   C_{\rm cor}(\gamma) = \frac{\exp\left[- \frac{2\tau}{3}f_0(\gamma)\right]}{
                         \left(1 + \left[\frac{\pi \tau}{6} 
                          f_2(\gamma)\right]^{6/7} \right)^{7/12}}  \;.
\label{Ccor}
\end{equation}
This correction significantly increases the range of validity of the 
quasistatic formula. Fig.~\ref{fig:Cor_QSFA} shows a direct comparison 
of $W_{\rm SFA}$ and $W^{\rm cor}_{\rm QSFA}$ (both scaled by $\omega^{-1}$) 
for those two case where QSFA failed most severely for the 1S state of 
a hydrogenlike atom, $\kappa=2.0\,$a.\,u.\ and $F=0.05$ or 0.1\,a.\,u.   
Besides the oscillatory behavior of SFA that is also not reproduced in the 
corrected QSFA, the overall agreement is very good, even if the rate 
varies by many orders of magnitude, if $\omega$ changes from 0 to 
0.1\,a.\,u. The results of the corrected QSFA are not only interesting 
for improving the QSFA, but they also confirm once more that the two 
steps made in the derivation of the QSFA are justified.  
Finally, it is worth noticing that the range of applicability
of Eq.(\ref{cQSFA}) is not restricted to a range of parameters that 
according to the Keldysh parameter $\gamma$ belongs to the quasistatic 
regime ($\gamma \ll 1$). Fig.~\ref{fig:Cor_QSFA} shows that it works 
even for $\gamma > 1$.

\section{\label{sec:conclusion}Conclusion}

The SFA (KFR theory in velocity gauge) was studied both 
analytically and numerically in the quasistatic limit. The derived analytical 
asymptotic expression (QSFA) shows that in the presence of long-range Coulomb 
interactions and thus for ionization of neutral or positively charged atoms 
or molecules the SFA rate is proportional to the laser frequency in this 
limit. This evidently unphysical result indicates a break-down of the SFA. 
Furthermore, this result shows that in contrast to the case of short-range 
potentials the SFA rate does not converge to the tunneling limit for weak 
fields, if long-range Coulomb interactions are present. The analytical 
result is supported by a numerical study for which an efficient scheme for 
the numerical evaluation of the SFA transition amplitude has been developed. 

Using different states of hydrogenlike atoms as an example, the predictions 
of the original SFA and the QSFA are compared to each other. It is found 
that QSFA allows for a rather accurate prediction of the SFA rate even 
for finite laser frequencies extending in some favorable cases to 
wavelengths of 500\,nm and below. It is shown that the validity regime 
of the QSFA can even be extended using a correction factor that is 
explicitly derived for 1S states. The large range of applicability 
of the QSFA is of practical interest, since its numerical evaluation 
is simpler than the one of the original SFA rate, especially in the 
IR and far-IR frequency regime. Furthermore, it is very convenient 
for studies of the frequency dependence of the SFA rate, since the 
QSFA is, besides a simple proportionality factor, $\omega$ independent. 
Thus the QSFA has to be evaluated for a given system and field strength 
only once. In turn, the relatively large range of laser frequencies 
in which QSFA and SFA agree demonstrates that also the SFA rate itself 
is in a wide range of laser parameters only proportional to $\omega$. 
An exception is the pronounced $\omega$ dependence due to channel 
closings that is not reproduced by QSFA.      

On the basis of a comparison of the QSFA result with the prediction of 
the Popov-Peremolov-Terent'ev (PPT) formula a Coulomb correction factor 
is derived. This factor is compared to a previously proposed one that 
was supposed to be successfully adopted in a wide range of calculations. 
It is discussed that part of this success may be due to the fact that the 
comparisons to experimental data was only possible on a relative scale. 
In this case a number of important terms missing in the previously proposed 
Coulomb correction factor is not visible. 

The goal of this work has been the derivation of an anlytical expression 
for the SFA in the quasistatic limit in the presence of long-range Coulomb 
interactions and a discussion of the resulting QSFA in comparison to SFA. 
The investigation of the validity of the SFA itself by comparing to the 
results of full solutions of the time-dependent Schr\"odinger equation 
is presently underway and will be discussed elsewhere.

\section*{Acknowledgments}
AS and YV acknowledge financial support by the {\it Deutsche 
Forschungsgemeinschaft}. AS is grateful 
to the {\it Stifterverband f\"ur die Deutsche Wissenschaft} (Programme  
{\it Forschungsdozenturen}) and the {\it Fonds der Chemischen Industrie} 
for financial support.

\appendix

\section{\label{app:Bcoef}  
Functions $B(\rho)$ for different states. }

In this Appendix function $B$ defined by Eq.\,(\ref{Bcoef}) is given 
explicitly for all states of hydrogen-like atoms fulfilling $n\leq 3$. 
(Note, for hydrogenlike atoms function $B$ is independent of $\kappa$.)  
\begin{eqnarray*}
B_{\rm 1S_0}(\rho)   &=& 1 \\
B_{\rm 2S_0}(\rho)   &=& 4 - 12 \rho^{-2} + 10 \rho^{-4} \\ 
B_{\rm 2P_0}(\rho)   &=& 2 \rho^{-2} \\ 
B_{\rm 2P_1}(\rho)   &=& 5 \rho^{-2} - 5 \rho^{-4} \\ 
B_{\rm 3S_0}(\rho)   &=& 9 - 72 \rho^{-2} + 220 \rho^{-4} 
                         - 280 \rho^{-6} + 126 \rho^{-8} \\
B_{\rm 3P_0}(\rho)   &=& 12 \rho^{-2} -30 \rho^{-4} + 21 \rho^{-6} \\
B_{\rm 3P_1}(\rho)   &=& 30 \rho^{-2} - 135 \rho^{-4} 
                         + \frac{399}{2}\rho^{-6}-\frac{189}{2}\rho^{-8}\\
B_{\rm 3D_0}(\rho)   &=& \frac{47}{4}\rho^{-4} - \frac{49}{2}\rho^{-6} 
                         + \frac{63}{4}\rho^{-8}  \\
B_{\rm 3D_1}(\rho)   &=& \frac{21}{2}\rho^{-4} - \frac{21}{2}\rho^{-6} \\
B_{\rm 3D_2}(\rho)   &=& \frac{189}{8}\rho^{-4} - \frac{189}{4}\rho^{-6} 
                         + \frac{189}{8}\rho^{-8}  
\end{eqnarray*}

\section{\label{app:Cor}  Correction for large $\tau$.}

In this Appendix the corrected QSFA given in Eq.\,(\ref{cQSFA}) is 
derived. Introducing $G=\gamma\rho$ and $\bar{v} = v/G$, we rewrite 
$i S(u_{+})$ in (\ref{iSup}) as 
\begin{equation}
   i S(u_{+}) = -\frac{\tau \rho^3}{3} f(G,\chi)
\end{equation}
where
\begin{equation}
   f(G,\chi)  = \int\limits_0^{-\chi+i} \frac{3}{2i} 
                \frac{(\bar{v} + \chi)^2 +1}{\sqrt{1 - G^2 \bar{v}^2}}\, 
                d\bar{v}   \; .
\end{equation}
The real part of $f(G,\chi)$ can be given using a Taylor expansion with 
respect to $\chi$ as 
\begin{equation}
   {\rm Re} f(G,\chi)  =  1 + f_0(G) + f_2(G) \chi^2 + \dots
\end{equation}
where
\begin{eqnarray*}
   f_0(G) &=& \frac{3(1+ 2 G^2) \sinh^{-1}G}{4 G^3} -  
              \frac{3\sqrt{1+G^2}}{4G^2} - 1 \\
   f_2(G) &=& \frac{3}{2}\left[ \frac{\sinh^{-1}G}{G} - 
              \frac{1}{\sqrt{1+G^2}}\right] \; .
\end{eqnarray*}
For small values of $G$ these functions can be well approximated by 
$f_0(G)\approx - G^2/10$, $f_2(G)\approx G^2/2$ and for $G<4$ they can 
be fitted with good accuracy by
\begin{eqnarray}
   f_0(G) &\approx& -\frac{11 G^2 (14 + 3 G^2)}{55(28+15 G^2) + 54 G^4} \\
   f_2(G) &\approx& \frac{ 5 G^2 + G^3}{10 + 9 G^2} e^{-G/5} \; .
\end{eqnarray}

A simple correction factor for the 1S state can now be obtained. Indeed, 
for this case the main contribution comes from $\rho \approx 1$ and one 
can multiply $W_{\rm QSFA}$ in Eq.~(\ref{W_QSFA}) by  
\begin{equation}
   C_{\rm cor} = \exp[- (2\tau/3) f_0(\gamma)] 
                 \exp[- (2\tau/3) f_2(\gamma) \chi^2  ]
\label{Cadd}
\end{equation}
The first term in (\ref{Cadd}) yields an exponential increase with 
$\gamma$. The second term introduces a damping for $\chi > \sqrt{d}$ 
where $d = (2\tau/3) f_2(\gamma)$. 
Using the approximate identity (valid within one percent) 
\begin{equation}
   \int\limits_{-\infty}^{\infty}  \frac{e^{-d \chi^2}}{ (1+ \chi^2)^2} d \chi 
      \approx
   \frac{\pi}{2} \left[1 + \left(\frac{\pi d}{4}\right)^{6/7}  \right]^{-7/12}
\end{equation}
to carry out the integration over $\chi$ one obtains the final result given 
in Eqs.\,(\ref{cQSFA}) and (\ref{Ccor}).


\bibliographystyle{apsrev} 

\begin{thebibliography}{13}
\expandafter\ifx\csname natexlab\endcsname\relax\def\natexlab#1{#1}\fi
\expandafter\ifx\csname bibnamefont\endcsname\relax
  \def\bibnamefont#1{#1}\fi
\expandafter\ifx\csname bibfnamefont\endcsname\relax
  \def\bibfnamefont#1{#1}\fi
\expandafter\ifx\csname citenamefont\endcsname\relax
  \def\citenamefont#1{#1}\fi
\expandafter\ifx\csname url\endcsname\relax
  \def\url#1{\texttt{#1}}\fi
\expandafter\ifx\csname urlprefix\endcsname\relax\def\urlprefix{URL }\fi
\providecommand{\bibinfo}[2]{#2}
\providecommand{\eprint}[2][]{\url{#2}}

\bibitem[{\citenamefont{Becker and Faisal}(2005)}]{sfa:beck05}
\bibinfo{author}{\bibfnamefont{A.}~\bibnamefont{Becker}} \bibnamefont{and}
  \bibinfo{author}{\bibfnamefont{F.~H.~M.} \bibnamefont{Faisal}},
  \bibinfo{journal}{J.\,Phys.\ B: At.\,Mol.\,Phys.}
  \textbf{\bibinfo{volume}{38}}, \bibinfo{pages}{R1} (\bibinfo{year}{2005}).

\bibitem[{\citenamefont{Becker et~al.}(2001)\citenamefont{Becker, Plaja,
  Moreno, Nurhuda, and Faisal}}]{sfa:beck01}
\bibinfo{author}{\bibfnamefont{A.}~\bibnamefont{Becker}},
  \bibinfo{author}{\bibfnamefont{L.}~\bibnamefont{Plaja}},
  \bibinfo{author}{\bibfnamefont{P.}~\bibnamefont{Moreno}},
  \bibinfo{author}{\bibfnamefont{M.}~\bibnamefont{Nurhuda}}, \bibnamefont{and}
  \bibinfo{author}{\bibfnamefont{F.~H.~M.} \bibnamefont{Faisal}},
  \bibinfo{journal}{Phys.\,Rev.\ A} \textbf{\bibinfo{volume}{64}},
  \bibinfo{pages}{023408} (\bibinfo{year}{2001}).

\bibitem[{\citenamefont{Kjeldsen and Madsen}(2006)}]{sfa:kjel06}
\bibinfo{author}{\bibfnamefont{T.~K.} \bibnamefont{Kjeldsen}} \bibnamefont{and}
  \bibinfo{author}{\bibfnamefont{L.~B.} \bibnamefont{Madsen}},
  \bibinfo{journal}{Phys.\,Rev.\ A} \textbf{\bibinfo{volume}{74}},
  \bibinfo{pages}{023407} (\bibinfo{year}{2006}).

\bibitem[{\citenamefont{Milo\v{s}evi\'{c}
  et~al.}(2006)\citenamefont{Milo\v{s}evi\'{c}, Paulus, Bauer, and
  Becker}}]{sfa:milo06}
\bibinfo{author}{\bibfnamefont{D.~B.} \bibnamefont{Milo\v{s}evi\'{c}}},
  \bibinfo{author}{\bibfnamefont{G.~G.} \bibnamefont{Paulus}},
  \bibinfo{author}{\bibfnamefont{D.}~\bibnamefont{Bauer}}, \bibnamefont{and}
  \bibinfo{author}{\bibfnamefont{W.}~\bibnamefont{Becker}},
  \bibinfo{journal}{J.\,Phys.\ B: At.\,Mol.\,Phys.}
  \textbf{\bibinfo{volume}{39}}, \bibinfo{pages}{R203} (\bibinfo{year}{2006}).

\bibitem[{\citenamefont{Keldysh}(1965)}]{sfa:keld65}
\bibinfo{author}{\bibfnamefont{L.~V.} \bibnamefont{Keldysh}},
  \bibinfo{journal}{Sov.\,Phys.\ JETP} \textbf{\bibinfo{volume}{20}},
  \bibinfo{pages}{1307} (\bibinfo{year}{1965}).

\bibitem[{\citenamefont{Reiss}(1980)}]{sfa:reis80}
\bibinfo{author}{\bibfnamefont{H.~R.} \bibnamefont{Reiss}},
  \bibinfo{journal}{Phys.\,Rev.\ A} \textbf{\bibinfo{volume}{22}},
  \bibinfo{pages}{1786} (\bibinfo{year}{1980}).

\bibitem[{\citenamefont{Faisal}(1973)}]{sfa:fais73}
\bibinfo{author}{\bibfnamefont{F.~H.~M.} \bibnamefont{Faisal}},
  \bibinfo{journal}{J.\,Phys.\ B: At.\,Mol.\,Phys.}
  \textbf{\bibinfo{volume}{6}}, \bibinfo{pages}{L89} (\bibinfo{year}{1973}).

\bibitem[{\citenamefont{Reiss}(1992)}]{sfa:reis92}
\bibinfo{author}{\bibfnamefont{H.~R.} \bibnamefont{Reiss}},
  \bibinfo{journal}{Prog.\,Quant.\,Electr.} \textbf{\bibinfo{volume}{16}},
  \bibinfo{pages}{1} (\bibinfo{year}{1992}).

\bibitem[{\citenamefont{Perelomov et~al.}(1966)\citenamefont{Perelomov, Popov,
  and Terent'ev}}]{sfa:pere66}
\bibinfo{author}{\bibfnamefont{A.~M.} \bibnamefont{Perelomov}},
  \bibinfo{author}{\bibfnamefont{V.~S.} \bibnamefont{Popov}}, \bibnamefont{and}
  \bibinfo{author}{\bibfnamefont{M.~V.} \bibnamefont{Terent'ev}},
  \bibinfo{journal}{Sov.\,Phys.\ JETP} \textbf{\bibinfo{volume}{23}},
  \bibinfo{pages}{924} (\bibinfo{year}{1966}).

\bibitem[{\citenamefont{Ammosov et~al.}(1986)\citenamefont{Ammosov, Delone, and
  Krainov}}]{sfa:ammo86}
\bibinfo{author}{\bibfnamefont{M.~V.} \bibnamefont{Ammosov}},
  \bibinfo{author}{\bibfnamefont{N.~B.} \bibnamefont{Delone}},
  \bibnamefont{and} \bibinfo{author}{\bibfnamefont{V.~P.}
  \bibnamefont{Krainov}}, \bibinfo{journal}{Sov.\,Phys.\ JETP}
  \textbf{\bibinfo{volume}{64}}, \bibinfo{pages}{1191} (\bibinfo{year}{1986}).

\bibitem[{\citenamefont{Reiss and Krainov}(2003)}]{sfa:reis03}
\bibinfo{author}{\bibfnamefont{H.~R.} \bibnamefont{Reiss}} \bibnamefont{and}
  \bibinfo{author}{\bibfnamefont{V.~P.} \bibnamefont{Krainov}},
  \bibinfo{journal}{J.\,Phys.\ A: Math.\, Gen.} \textbf{\bibinfo{volume}{36}},
  \bibinfo{pages}{5575} (\bibinfo{year}{2003}).

\bibitem[{\citenamefont{Bauer}(2005)}]{sfa:baue05c}
\bibinfo{author}{\bibfnamefont{J.}~\bibnamefont{Bauer}},
  \bibinfo{journal}{J.\,Phys.\ A: Math.\, Gen.} \textbf{\bibinfo{volume}{38}},
  \bibinfo{pages}{521} (\bibinfo{year}{2005}).

\bibitem[{\citenamefont{Bauer}(2006)}]{sfa:baue06}
\bibinfo{author}{\bibfnamefont{J.}~\bibnamefont{Bauer}},
  \bibinfo{journal}{Phys.\,Rev.\ A} \textbf{\bibinfo{volume}{73}},
  \bibinfo{pages}{023421} (\bibinfo{year}{2006}).

\end{thebibliography}

\end{document}